\documentclass[twocolumn,aip,jcp,lengthcheck,showpacs]{revtex4-1}
\usepackage[]{amsmath}
\usepackage{amssymb}
\usepackage{mhchem}
\usepackage{siunitx}
\usepackage{graphicx}
\usepackage{bm}
\usepackage{times,psfrag,subfigure}
\usepackage[]{color}
\usepackage{lmodern}
\usepackage{hyperref}
\begin{document}

\title{Morphological Transitions during Melting of Small Cylindrical Aggregates}
\thanks{This document is the unedited author's version of a Submitted Work that was subsequently accepted for publication and \textit{J. Phys. Chem. C}, copyright\textcopyright American Chemical Society after peer review. To access the final edited and published work see DOI: \url{10.1021/acs.jpcc.6b05434}}
\author{Chenyu Jin}
\email{chenyu.jin@ds.mpg.de}
\altaffiliation[Present address: ]{Max Planck Institute of Dynamics and Self-Organisation, Am Fa\ss berg 17, 37077 G\"ottingen, Germany}
\author{Hans Riegler}
\email{Hans.Riegler@mpikg.mpg.de}
\thanks{Corresponding author}
\affiliation{Max Planck Institute of Colloids and Interfaces, Science Park Golm, 14476 Potsdam, Germany}
\date{\today}

\begin{abstract}
  % background 1-2
Most studies on melting under confinement focus only on the solid and liquid melt phases. 
Despite of its ubiquity, contributions from the capillary interface (liquid / vapor interface) are often neglected.
  % methodology 2-3
In this study the melting behavior of small cylindrical aggregates in vapor attached to planar surfaces is analyzed. 
For the assumed boundary conditions (cylindrical solid with a non wetting top plane and a wettable side wall) solid and the liquid phases can coexist within a certain temperature range.
Due to capillary instability, the liquid phase can form either an axisymmetric rouloid morphology or, above a certain threshold liquid volume fraction, a bulge coexisting with a rouloid-like section.
The corresponding melting points are different.
% There can be an energy barrier between the two morphologies.
% The corresponding melting points are different and can be above or below the bulk melting point depending on the size and shape of the aggregates.  
% The axisymmetric rouloid configuration is metastable up to a certain temperature.
% Bulged morphologies are always unstable and melt instantaneously.
  % conclusion and highlight 1
The analysis explicitly describes the behavior of a real system of small aggregates of long chain alkanes on planar substrates.
It also gives qualitative insights into the melting behavior of small aggregates with anisotropic wetting behaviors in general.
It reveals in particular how melting points and melting pathways depend on the energetic respectively morphological pathways leading to complete melting.
\end{abstract}
 
\maketitle

\section{Introduction}

The thermodynamics of reduced dimensions (small particles, confined materials, etc.) is of interest in fundamental and applied research. 
For instance, the melting behavior depends on the system size and shape \cite{alba2006effects,christenson2001confine, gubbins2014thermodynamics}, resulting from the competition between the changes of the bulk energies and the changes of the interfacial energies that evolve during the phase transition.

In simple, highly symmetrical cases with isotropic interfacial energies the phase transition can be described by a Gibbs-Thomson effect (GT) \cite{Pawlow1909a, tammann1920methode, meissner1920, rie1923}, which is widely used by experimentalists and engineers \cite{Takagi1954TEM, Borel1976goldED, Couchman1977nature, Reiss1988coexist, LuKe2007rev, morishige2010effect}.  
A similar approach has been used to describe nucleation phenomena or the solubility of small particles \cite{skinner1972kelvin, fu2013confined, huber2015soft}.
Its extension, the Gibbs-Thomson-Herring approach (GTH) takes into account anisotropic interfacial energies between mother and daughter phase \cite{herring1951some, nozieres1992shape}.
Thus it can be applied for instance to analyze the melting of small faceted solid aggregates embedded in an infinite bulk volume of its melt.

Rarely analyzed yet is the melting behavior of small anisotropic systems including a capillary interface i.e., the liquid/vapor interface.
In this case energetic contributions from \emph{two} interfaces affect the melting behavior.
During melting the area and the geometry of the solid/liquid interface, as well as the ``capillary'' interface between the two fluids (liquid and vapor) evolve on a path given by the minimization of the interfacial and the bulk energies of the entire system. 

Here we analyze the melting behavior of small cylindrical aggregates in a gaseous environment.
We show that their melting scenarios include morphological transitions of the capillary interface, which have an impact on the melting paths and on the melting points. 
The analysis is motivated by experiments with cylinder-like aggregates of long chain alkanes on planar interfaces \cite{Lazar2005movingdrop} and by an earlier theoretical analysis on straight terraces \cite{Halim2012bulge}.
Cylindrical aggregates can be found in micro- and nano-manufacturing, e.~g., the epitaxial growth of nano-wires \cite{wang2008growth}, or the 3D printing of small structures \cite{visser2015toward}.
Our basic assumption that the solid core always retains the shape of a right cylinder may hold only for a few real systems.
Yet, the aim of this article is to explore the general aspects of the melting behavior of small, anisotropic systems, with a focus on the impact of the morphology of the capillary interface on the melting behavior.

\section{Theory and simulation details}
\subsection{General theoretical approach}
We analyze the solid/liquid phase transition in a continuum approach scenario\footnote{B. Henrich et al. showed that continuum models are adequate for $\geqslant 2$ adjacent fluid mono-layers using computer simulation \cite{henrich2008continuum}.}.
Contributions from line tension \cite{tolman1949effect}, disjoining pressure \cite{derjaguin1978} and gravity are neglected.
The total free energy of the system, $G_\text{total}=G_\text{B}+G_\text{I}$, is the sum of energy contributions from the bulk, $G_\text{B}$ and from all interfaces, $G_\text{I}$.
The chemical potential per unit volume of the liquid (l) and solid (s) phases 
\footnote{liquid, liquid melt, or melt are used here as synonyms for the liquid (molten) phase.}
at temperature $T$ and pressure $p$ is $dG_\text{B}/dV=\Delta \mu $ with $\Delta \mu = \mu_\text{l}(T,p) -\mu_\text{s}(T,p)$.

The system is assumed in thermodynamic equilibrium with $dG_\text{total}/dV_\text{l}=0$:
\begin{equation}
  \Delta \mu +\frac{dG_\text{I}}{dV_\text{l}} =0
  \label{eqn:equil}
\end{equation}

$\Delta \mu  \simeq -S_\text{m}\cdot (T-T_\text{0})$ is approximated with a constant melting entropy $S_\text{m}$ (the relevant temperature range is very small)  \cite{adkins1983equilTD, Dash2006premeltice}.
$T_\text{0}$ is the bulk melting temperature.
This approximation of $\Delta \mu$ together with eq. (\ref{eqn:equil}) defines an "equilibrium" temperature if $dG_\text{I}/dV_\text{l} < 0$.
In this case the system is in a metastable state with a certain fixed amount of solid and liquid coexisting:
\begin{equation}
  \Delta T_\text{eq}=T_\text{eq} - T_\text{0} = \frac{1}{S_\text{m}}\cdot (\frac{dG_\text{I}}{dV_\text{l}})
  \label{eqn:T}
\end{equation}

The upper limit of $T_\text{eq}$ i.e., the maximum change of the interfacial energy caused by an infinitesimal volume change, $({dG_\text{I}}/{dV_\text{l}})_\text{max}$, is  

\begin{equation}
 \Delta T_\text{m}=T_\text{m} - T_\text{0} =\frac{1}{S_\text{m}}\cdot (\frac{dG_\text{I}}{dV_\text{l}})_\text{max}
  \label{eqn:Tm}
\end{equation}

This $T_\text{m}$ presents the highest possible equilibrium temperature. 
At $T_\text{m}$ the system can start to melt completely without having to overcome an energy barrier.
$T_\text{m}$ is the ``de facto''%%%footnote
\footnote{also called the ``true'' or the ``critical'' melting temperature. We use the term ``critical'' for another specific temperature (see below), hence the term ``de facto''.} melting temperature of the system.

\begin{figure} 
  \centering
  \includegraphics[width=\columnwidth]{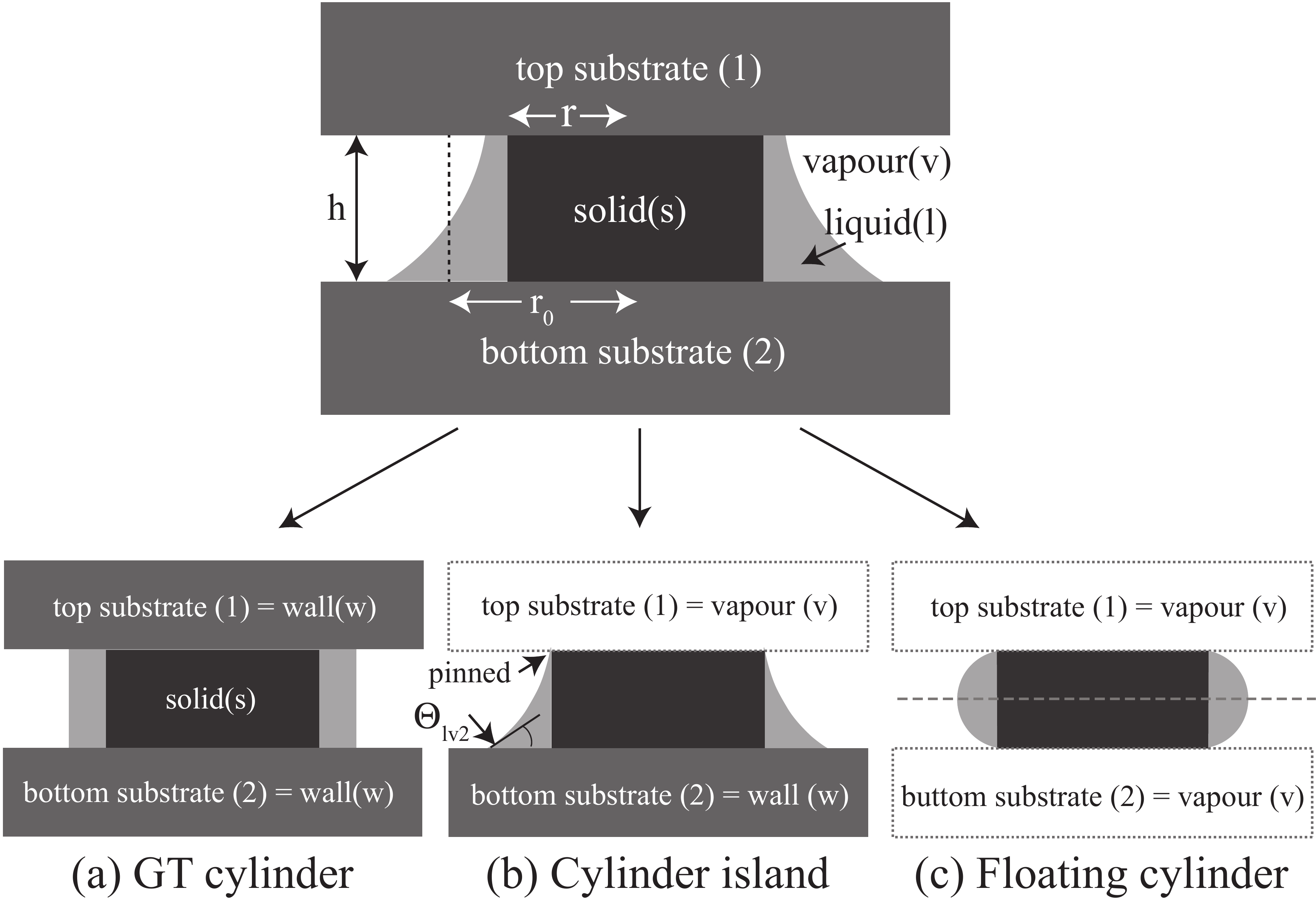}
  \caption{Cross-section through a solid core of right cylindrical geometry sandwiched between two parallel substrates.
  The cylinder sidewall is completely wetted by liquid melt.
 The different cartoons show the result of various wettabilities of the substrate surfaces by the melt.
  }
  \label{fig:cylinders}
\end{figure}

Let us analyze the melting behavior of aggregates with cylindrical shapes as depicted in fig.~\ref{fig:cylinders}.
We assume that: 

1.) the solid maintains a smooth, right  cylinder geometry with a constant height $h$; 

2.) the total volume of solid and liquid phase together remains constant%%%footnote
\footnote{typically the density of solid and liquid is quite similar  \cite{seyer1944density}.}; 

3.) the substrates are planar;

4.) the cylinder side walls are completely wetted by the liquid; 

5.) the top and bottom base planes of the solid cylinder are not wetted by the liquid;

6.) Young's equation holds for contact angles at the liquid/substrate/vapor contact lines.

The upper schematic of fig.~\ref{fig:cylinders} shows such a system in the situation of solid/liquid coexistence for the most general case in agreement with these assumptions. 
The three schematics below show more specific cases.
In the all-solid state the round top and bottom planes of the solid cylinders have radii $r_0$.
If some solid has been transferred into the liquid state the radius of the solid cylinder core is $r$.

In case (a) the contact angles of the solid and the liquid phase with the substrates are exactly $90^{\circ}$. 
This case represents a Gibbs-Thomson ("GT cylinder") approximation for the melting of a cylindrical solid aggregate i. e., the melting is affected only by the competition between the energies of the bulk phases and of the solid/liquid interface \cite{Pawlow1909a, tammann1920methode, meissner1920, rie1923}. 
The melting is not influenced by the contact of the liquid and solid with the top and bottom substrate, it is independent from $h$.

Cases (b) ("Cylinder island") and (c) ("Floating cylinder") assume that the liquid is "pinned" at the solid/vapor/liquid contact line of the top cylinder plane.
Pinning means that there is no fixed contact angle at this line. 
The contact angle adjusts itself such that the sum of the interfacial energies of the interfaces (i. e.,  their geometry), is minimised.
In case (b) the cylinder is deposited on a substrate.
In case (c) top and bottom planes of the solid cylinder are both in contact with the surrounding vapor phase.
Case (c) represents a special case (b).
A floating cylinder can be divided into two identical cylinder islands (dividing plane indicated by the dashed line). 
Each of these two cylinder islands has on one base plane side a contact angle of $\Theta_\text{lv2}=90^\circ$.
On the other base plane side the liquid is pinned at the base plane perimeter.
Case (c) with height $h$ is identical to case (b) with height $h/2$ and $\Theta_\text{lv2}=90^\circ$. 
 
In the most general cylinder case (upper schematic of fig.~\ref{fig:cylinders}), the sum of all interfacial energies is:
\begin{align}
 G_\text{I}&=(\gamma_\text{s1}-\gamma_\text{v1})A_\text{s1}+(\gamma_\text{s2}-\gamma_\text{v2})A_\text{s2} \nonumber \\
 &+(\gamma_\text{l1}-\gamma_\text{v1})A_\text{l1}+(\gamma_\text{l2}-\gamma_\text{v2})A_\text{l2}+\gamma_\text{ls}A_\text{ls}+\gamma_\text{lv}A_\text{lv}
\label{eqn:A1}
\end{align}

With assumption \#1 ($A_\text{s1}$ = $A_\text{s2}=\pi r^2$) this means:
\begin{align}
 G_\text{I}&=(\gamma_\text{s1}-\gamma_\text{v1}+\gamma_\text{s2}-\gamma_\text{v2})\pi r^2\nonumber \\
&+(\gamma_\text{l1}-\gamma_\text{v1})A_\text{l1}+(\gamma_\text{l2}-\gamma_\text{v2})A_\text{l2}+\gamma_\text{ls}A_\text{ls}+\gamma_\text{lv}A_\text{lv}\nonumber \\
 G_\text{I}&=(\gamma_\text{s1}-\gamma_\text{v1}+\gamma_\text{s2}-\gamma_\text{v2})\pi r^2\nonumber \\
&+ G_\text{I}(liquid)
\label{eqn:A2}
\end{align}

$G_\text{I}(liquid)$ summarises all the energy contributions from interfaces of the liquid phase with its environments. 
Assumption \#2 ($V_\text{l} = \pi h(r_\text{0}^2-r^2)$) for eq.~(\ref{eqn:A2}) yields:
\begin{equation}
  \frac{dG_\text{I}}{dV_\text{l}} = \frac{dG_\text{I}(liquid)}{dV_\text{l}} -\frac{1}{h} (\gamma_\text{s1}-\gamma_\text{v1}+\gamma_\text{s2}-\gamma_\text{v2})
  \label{eqn:A3}
\end{equation} 

And with eq.~(\ref{eqn:T}) we obtain:
\begin{equation}
 \Delta T_\text{eq} = \frac{1}{S_\text{m}} \left[ (\frac{dG_\text{I}(liquid)}{dV_\text{l}})-\frac{
     (\gamma_\text{s1}-\gamma_\text{v1}+\gamma_\text{s2}-\gamma_\text{v2})}{h} \right]
      \label{eqn:A4}
\end{equation} 

With eq.~(\ref{eqn:A4}) the shift of the equilibrium temperature (and also of the ``de facto'' melting temperature, $T_\text{m}$) is determined by (a) the (geometry) evolution of the interface(s) between the liquid phase and its various environments (first term on the right side) and (b) an offset given by the interfacial energies of the solid top and bottom planes scaled to the height of the cylinder (second term). 

In the following we will focus on the behavior of a cylinder island as depicted in fig.~\ref{fig:cylinders}(b) (see Appendix for cases (a) and (c)). 
For the cylinder island eq.~(\ref{eqn:A3}) reads as:
\begin{equation}
  \frac{dG_\text{I}^\text{island}}{dV_\text{l}} = \frac{dG_\text{I}(liquid)}{dV_\text{l}} -\frac{1}{h} (\gamma_\text{sv}+\gamma_\text{sw}-\gamma_\text{vw})
  \label{eqn:A7}
\end{equation} 

and thus:
\begin{align}
  \Delta T_{eq}^\text{island}= \frac{1}{S_\text{m}} \left[ (\frac{dG_\text{I}(liquid)}{dV_\text{l}})-\frac{(\gamma_\text{sv}+\gamma_\text{sw}-\gamma_\text{vw})}{h}\right]
  \label{eqn:A8}
\end{align}

%%%%%%%%%%

\subsection{The basic metastable solid/liquid coexistence configurations: Rouloids and bulged morphologies}

\begin{figure}
	\centering
	\includegraphics[width=\columnwidth]{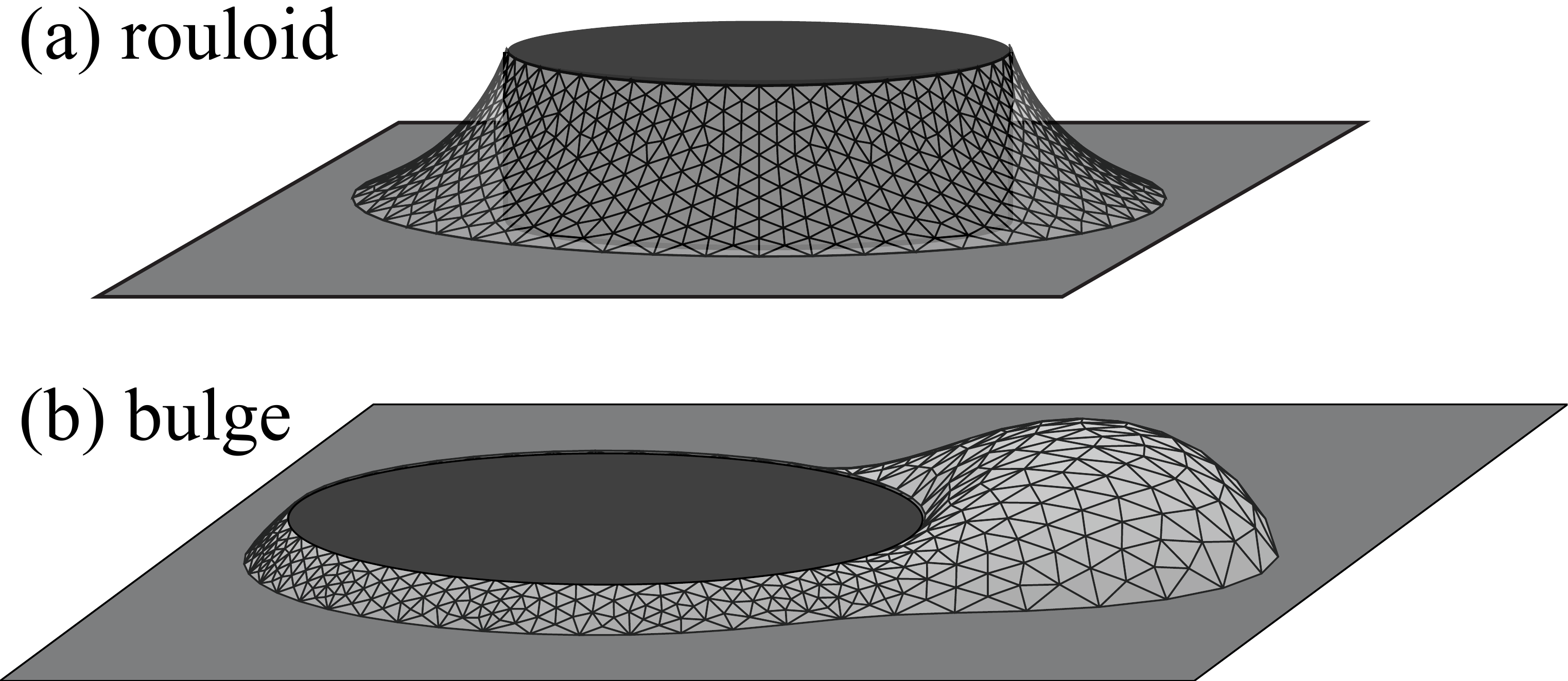}
	\caption{The two possible morphologies of the liquid/vapor interface.
	It is assumed that the solid retains a right cylinder shape and that the liquid wets the substrates as depicted in fig.~\ref{fig:cylinders}(b) ("cylinder islands").   
	}
	\label{fig:island-scheme}
\end{figure}

Eq~(\ref{eqn:A8}) was solved by numerical simulations with Surface Evolver \cite{brakke1992evolver} for various cylinder shapes $r_\text{0}/h$.
For the simulations real data of typical long-chain alkane systems are assumed with $\gamma_\text{lv}=25 \times 10^{-3}$N/m, $\gamma_\text{lw}=4 \times 10^{-3}$N/m, $\gamma_\text{ls} \text{(cylinder side wall)} = 10 \times 10^{-3}$N/m, $\gamma_\text{ls}\text{(cylinder base plane)} = 4 \times 10^{-3}$N/m,  and $\Delta S = q_\text{m}/T_\text{m} = 5 \times 10^5$ J/Km$^{-3}$ \cite{Dirand2002enthalpy, yi2011MD, riegler2007pre}.
The contact angle is $\theta_\text{lvw}=15^\circ$.
The results can be compared to the observed melting scenarios of round multilayer islands of long chain alkanes on planar surfaces \cite{Merkl1997SF, holzwarth2000molecular, Lazar2005flow} (see Appendix).

The simulations reveal that liquid and solid can coexist in a metastable state within a certain temperature range.  
Within this liquid/solid coexistence regime, for certain temperature ranges, two distinctly different, isochoric types of morphologies are possible (fig.~\ref{fig:island-scheme}).
Both morphologies have isocurvature liquid/vapor interfaces, assuring isobaric conditions (mechanical equilibrium) for the liquid phase.    
One type of morphology are rouloids (axisymmetric channels) as studied by Delaunay \cite{delaunay1841surface, Eells1987surfaces} (fig.~\ref{fig:island-scheme}(a).
Such rouloids are the typical shapes of capillary bridges of liquids confined between planar substrates \cite{heady1970analysis, carter1988forces, langbein2002capillary}. 
The other type of morphology are bulged liquid/vapor interface sections in isobaric (isocurvature) coexistence with rouloid-like channel sections (fig.~\ref{fig:island-scheme}(b).
Such bulges have been reported before in the case of wetting of planar patterned surfaces by liquid stripes \cite{gau1999liquid, lenz2001wetting, berthier2012physics} and for the melting behavior of  terraces with straight edges \cite{Halim2012bulge}.
There are temperature ranges, where rouloids or bulged morphologies are possible and there are ranges (below $T_\text{m}$), where only rouloids are possible.
Which morphology is possible depends on 

1.) $r_\text{0}/h$, the shape of the cylinder in its all-solid state

2.) the (non) wetting properties of the substrate, and

3.) the volume fraction of the liquid melt.

\section{Result and Discussion}
\subsection{The melting scenarios} 
 
\begin{figure}
  \centering
  \includegraphics[width=\columnwidth]{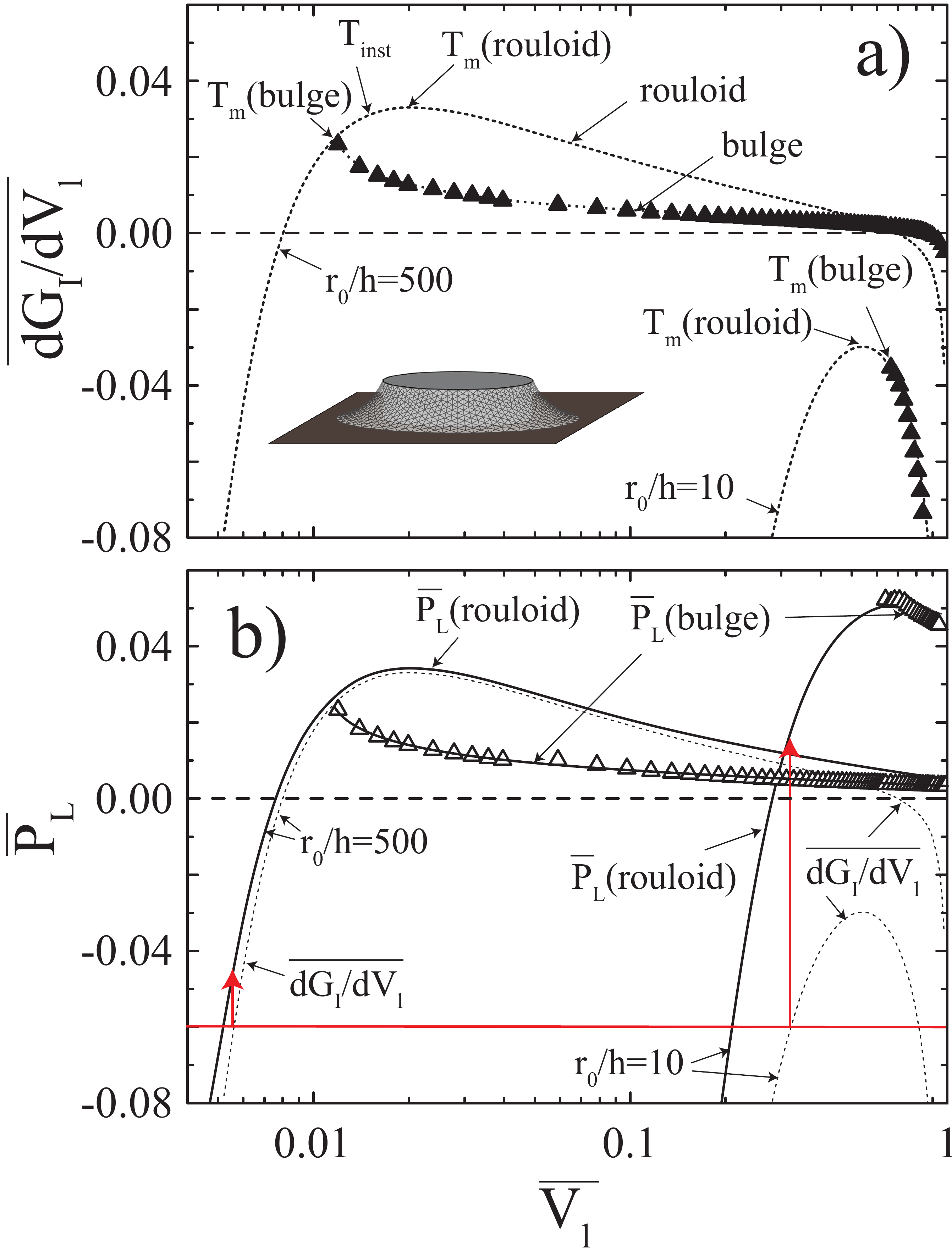}
  \caption{Melting scenarios of cylindrical islands on a planar substrate with wetting conditions as in fig.~\ref{fig:cylinders}(b).
  The data result from simulations assuming the thermodynamic data of islands of long-chain alkanes.
  (a) $\overline{dG_\text{I}/dV_\text{l}}(=dG_\text{I}/dV_\text{l}\cdot h/\gamma_\text{lv})$ as a function of the liquid volume fraction in the total volume, $\overline{V_\text{l}}(=V_\text{l}/\pi r^2_0 h$). 
  The behavior of the rouloids and the bulged shape ($\overline{dG_\text{I}/dV_\text{l}}$) are depicted by the dotted lines and filled triangles respectively.
  Please note that $\overline{dG_\text{I}/dV_\text{l}} \propto T_{eq}-T_0$.
  $T_\text{m}\text{(rouloid)}$ and $T_\text{m}\text{(bulge)}$ indicate the de facto melting temperatures of the corresponding morphology.
 At  $T_{inst}$ rouloids can form bulged shape without energy barrier (see fig.~\ref{fig:transitions}).
 (b) $\overline{P_\text{L}}(=P_\text{L}\cdot h/\gamma_\text{lv}$) as a function of the liquid volume fraction for rouloids(solid lines) and the bulged morphology (empty triangles).
% The dotted lines show the corresponding $\overline{dG_\text{I}/dV}$ as in (a). 
The endpoint of the red arrows indicate the isothernal Laplace pressures for the two island shapes for rouloid geometries (for $T$ corresponding to $\overline{dG_\text{I}/dV}=-0.06$).
 }
  \label{fig:equil-dG}
\end{figure}

Fig.~\ref{fig:equil-dG}(a) shows $\overline{dG_\text{I}/dV_\text{l}}=dG_\text{I}/dV_\text{l} \cdot h/\gamma_\text{lv}$ as function of the liquid volume fraction in the total material volume, $\overline{V_\text{l}}$ ($=V_\text{l}/\pi r^2_0 h$).
Because $\overline{dG_\text{I}/dV_\text{l}} \propto T_{eq}-T_0$ (Eq.~\eqref{eqn:T}), the plot explicitly reveals the difference between the (metastable) equilibrium temperatures of the system and the bulk melting temperature.
The bulk melting point, $T_\text{0}$, is at $\overline{dG_\text{I}/dV_\text{l}}=0$.
Data are shown for "low" ($r_\text{0}/h=500$) and "tall" ($r_\text{0}/h=10$) islands for rouloid and for bulged morphologies.
For the rouloid morphology, for both island shapes, starting from well below $0$, $\overline{dG_\text{I}/dV}$ first increases steeply with increasing $\overline{V_\text{I}}$.
Eventually $\overline{dG_\text{I}/dV}$ reaches  a maximum.
This corresponds to $T_\text{m}\text{(rouloid)}$ (Eq.~\eqref{eqn:Tm}).
Beyond $T_\text{m}\text{(rouloid)}$ $\overline{dG_\text{I}/dV}$ continuously decreases with increasing $\overline{V_\text{I}}$. 
For the bulged morphology $\overline{dG_\text{I}/dV_\text{l}}$  is depicted by the lines connecting the (simulation) data points of the full triangles.
In contrast to $\overline{dG_\text{I}/dV}\text{(rouloid)}$, there is a non-zero lower volume limit for $\overline{dG_\text{I}/dV}\text{(bulge)}$.
Below this limiting volume bulged morphologies are not possible (because there is not enough liquid to form a bulge in coexistence with a channel section with the same surface curvature).
As $\overline{dG_\text{I}/dV_\text{l}}\text{(bulge)}$ decreases monotonically, this lower volume limit also corresponds to the de facto melting points, $T_\text{m}\text{(bulge)}$. 
For $r_\text{0}/h=500$, $T_\text{m}\text{(bulge)}$ is smaller than $T_\text{m}\text{(rouloid)}$.
Also, $\overline{dG_\text{I}/dV_\text{l}}\text{(bulge)}$ remains smaller than $\overline{dG_\text{I}/dV_\text{l}}\text{(rouloid)}$ between $T_\text{m}\text{(bulge)}$ and a cross-over point at rather large values of $\overline{V_\text{I}}$.
% For very large  $\overline{V_\text{I}}$, $\overline{dG_\text{I}/dV_\text{l}}\text{(rouloid)}$ becomes smaller than $\overline{dG_\text{I}/dV_\text{l}}\text{(bulge)}$. 
$T_\text{m}\text{(rouloid)}$ and $T_\text{m}\text{(bulge)}$ are both above the bulk melting point, $T_\text{0}$.

For the tall island with $r_\text{0}/h=10$, $\overline{dG_\text{I}/dV}\text{(bulge)}$ also has a lower volume limit, indicated by $T_\text{m}\text{(bulge)}$.
This $T_\text{m}\text{(bulge)}$ is not a melting temperature, because it is on the downhill branch of $\overline{dG_\text{I}/dV}\text{(rouloid)}$ and the island is already melting irreversibly in a rouloid shape.

Fig.~\ref{fig:equil-dG}(b) shows a scaled Laplace pressure, $\overline{P_\text{L}}=P_\text{L}\cdot h/\gamma_\text{lv}$, as function of $\overline{V_\text{l}}$ ($=V_\text{l}/\pi r^2_0 h$) and again $\overline{dG_\text{I}/dV_\text{l}}$ as in Fig.~\ref{fig:equil-dG}(a).
$\overline{P_\text{L}}$ is depicted for low and tall islands, for rouloids and bulges. 
For the low islands with $r_\text{0}/h=500$, with the selected scalings of $P_\text{L}$ and $dG_\text{I}/dV$, the behavior of $\overline{P_\text{L}}$ and $\overline{dG_\text{I}/dV}$ is qualitatively and quantitatively pretty much identical  for both morphologies.
For the taller islands, in contrast, with $r_\text{0}/h=10$, $\overline{P_\text{L}}$ and $\overline{dG_\text{I}/dV}$ are significantly different.
$\overline{P_\text{L}}$  can reaching quite high positive values at $T_\text{m}\text{(bulge)}$.
The maximum values of $\overline{P_\text{L}}$ of the tall islands are higher than those of the low islands.

\begin{figure}
  \centering
  \includegraphics[width=\columnwidth]{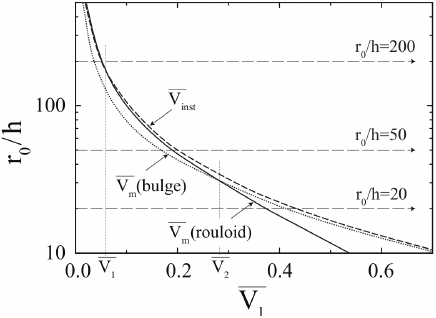}
  \caption{Liquid volume fractions of the rouloid morphology ($\overline{V_\text{m}}\text{(rouloid)}$) and of the bulged morphology ($\overline{V_\text{m}}\text{(bulge)}$) as function of the shape, $r_\text{0}/h$, and of the liquid volume fraction in the total volume, $\overline{V_\text{l}}(=V_\text{l}/ r^2_0 \pi$).
  Also depicted is the scaled liquid volume fraction, $\overline{V_\text{inst}}$, where the energy barrier between the rouloid and the bulged morphology vanishes.
   }
  \label{fig:transitions}
\end{figure}

Fig.~\ref{fig:transitions} presents as ordinate the shape of the cylinders, $r_0/h$, versus the scaled liquid volume fraction, $\overline{V_\text{l}}$, as abscissa.
The curves in the plot show the scaled liquid volume fractions at the de facto melting points of the rouloid ($\overline{V_\text{m}}\text{(rouloid)}$) and of the bulged morphologies ($\overline{V_\text{m}}\text{(bulge)}$). 
Also depicted is the scaled volume fraction where an isochoric transition between both morphologies is possible without energy barrier ($\overline{V_\text{inst}}$)%%%footnote%%%
\footnote{$\overline{V_\text{inst}}$ corresponds to a temperature $T_\text{inst}$ (fig.~\ref{fig:equil-dG}(a)) for the island with $r_\text{0}/h=500$).   
In addition, $\overline{V_\text{1}}$ marks the volume with $\overline{V_\text{inst}}= \overline{V_\text{m}}\text{(rouloid)}$ and $\overline{V_\text{2}}$ the volume, where $\overline{V_\text{inst}}= \overline{V_\text{m}}\text{(bulge)}$.}.%%%%

Figures~\ref{fig:equil-dG}~and~\ref{fig:transitions} reveal the various melting scenarios of low ($r_\text{0}/h=500$) and tall ($r_\text{0}/h=10$) islands assuming that the systems are in their energy minimum according to Eq.~\eqref{eqn:T}. 
Fig.~\ref{fig:equil-dG}(a) shows that:

1.) For the rouloid morphology, for $r_\text{0}/h<\infty$, in the ideal continuum approach, $\overline{dG_\text{I}/dV_\text{l}}\text{(rouloid)}$ always has an uphill and a downhill branch with a maximum at $T_\text{m}\text{(rouloid)}$. 
The downhill branch shows system states that are absolutely unstable.
The system will instantaneously starts its complete melting process on a rouloid pathway %%%footnote%%%
\footnote{here and in the following scenarios we only focus on the liquid channel morphology at the very \emph{start} of the complete melting process. The morphology at later stages or morphological transitions that occur when the complete melting advances are not considered here.}%%%
. 
On the uphill branch the system is metastable.
There are always two isothermal $\overline{V_\text{l}}$, with the larger $\overline{V_\text{l}}$ on the unstable downhill section of the $\overline{dG_\text{I}/dV_\text{l}}\text{(rouloid)}$-curve.

2.) For the rouloid morphology there is always an energetic barrier for an isothermal transition between uphill and downhill states.

3.) For small $\overline{V_\text{l}}$ and well below $T_\text{0}$,  the system is always in the rouloid configuration on the uphill branch of the $\overline{dG_\text{I}/dV_\text{l}}\text{(rouloid)}$-curve. 

4.) $\overline{dG_\text{I}/dV_\text{l}}\text{(bulge)}$ always starts at a non-zero lower limit volume.
It starts with its maximum value  $(\overline{dG_\text{I}/dV_\text{l}}\text{(bulge)})_\text{max}$ i. e., at $\overline{T_\text{m}}\text{(bulge)}$.
$\overline{dG_\text{I}/dV_\text{l}}\text{(bulge)}$ always goes downhill.
Therefore all bulged morphologies are unstable in this melting scenario.

5.) For large $r_\text{0}/h$, $T_\text{m}\text{(bulge)}$ is lower than $T_\text{m}\text{(rouloid)}$. 
For small $r_\text{0}/h$  the melting point $T_\text{m}\text{(bulge)}$ does not exist.
This is reflected in the behavior of the corresponding melting volumes, $\overline{V_\text{m}}\text{(bulge)}$ and $\overline{V_\text{m}}\text{(rouloid)}$, as depicted in Fig.~\ref{fig:transitions}.

6.) There can be energetic barriers for the isothermal transition between the rouloid and the bulged morphology.
This barrier disappears at $\overline{V_\text{inst}}$ corresponding to a certain $T_\text{inst}$.
For large $r_\text{0}/h$, $V_\text{inst}$, is below $V_\text{m}\text{(rouloid)}$, for low $r_\text{0}/h$, $V_\text{inst}$, is above $V_\text{m}\text{(rouloid)}$.

If we ignore fluctuations that overcome energy barriers, we obtain the following heating/melting scenarios.
We can heat the low island ($r_\text{0}/h=500$) up to $T_\text{inst}$.
This temperature is above the bulk melting point $T_\text{0}$ and above the melting point of the bulged morphology, $T_\text{m}\text{(bulge)}$, but below the melting point of the rouloid morphology, $T_\text{m}\text{(rouloid)}$.
Upon heating to $T_\text{inst}$, the increasing liquid fraction will retain a rouloid morphology. 
At $T_\text{inst}$ this rouloid morphology will become unstable and transform isochorically into a bulged morphology.
Then the system will instantaneously start to melt completely.
The taller island with $r_\text{0}/h=10$ can be heated to $T_\text{m}\text{(rouloid)}$, which is in this case below $T_\text{0}$.
At $T_\text{m}\text{(rouloid)}$ it will commence to melt completely, starting on a rouloid pathway.

The red lines in Fig.~\ref{fig:equil-dG}(b) reveal the differences in the Laplace pressures $\overline{P_\text{L}}=P_\text{L}\cdot\text{h}/\gamma_\text{lv}$ between the two islands with $r_\text{0}/h=10$ and $r_\text{0}/h=500$ under isothermal conditions. 
For instance, at a temperature corresponding to $\overline{dG_\text{I}/dV} = 0.06$ is the scaled Laplace pressure much higher for the tall island ($\overline{P_\text{L}}\approx 0.01$) than for the low island ($\overline{P_\text{l}}\approx -0.04$)%%footnote%%%
\footnote{the Laplace pressure can be negative and still some solid can be molten even below the bulk melting point if the liquid wets some of the substrate surface (if $\Theta < 90^\circ$).}%%%
.
At this temperature both islands are metastable against complete melting.
If two islands of the same height $h$ but different lateral size $r_0$ are connected isothermally by a (precursor) film of mobile molecules, this difference  in $\overline{P_\text{L}}$ will drive a liquid transport from the smaller to the larger island and the larger island will grow at the expense of the smaller one.
Such an Ostwald ripening process indeed has been reported for a system with (cylindrical) islands of long-chain alkanes at planar substrates \cite{pithan2015}.
It has to be noted that islands of identical shape, $r_\text{0}/h$, do have identical $\overline{P_\text{L}}$. 
However, they do have different Laplace pressures $P_\text{L}$, island with a smaller $h$ has the higher $P_\text{L}$.
Because $\overline{P_\text{L}}$ is a function of $r_\text{0}$, $h$, $\gamma_\text{lv}$, $\cos \Theta_\text{lv}$, and other system parameters, "larger" islands do not necessarily have a lower $P_\text{L}$ than "smaller" islands.

\subsection{Melting point shifts} 

\begin{figure}
  \centering
  \includegraphics[width=\columnwidth]{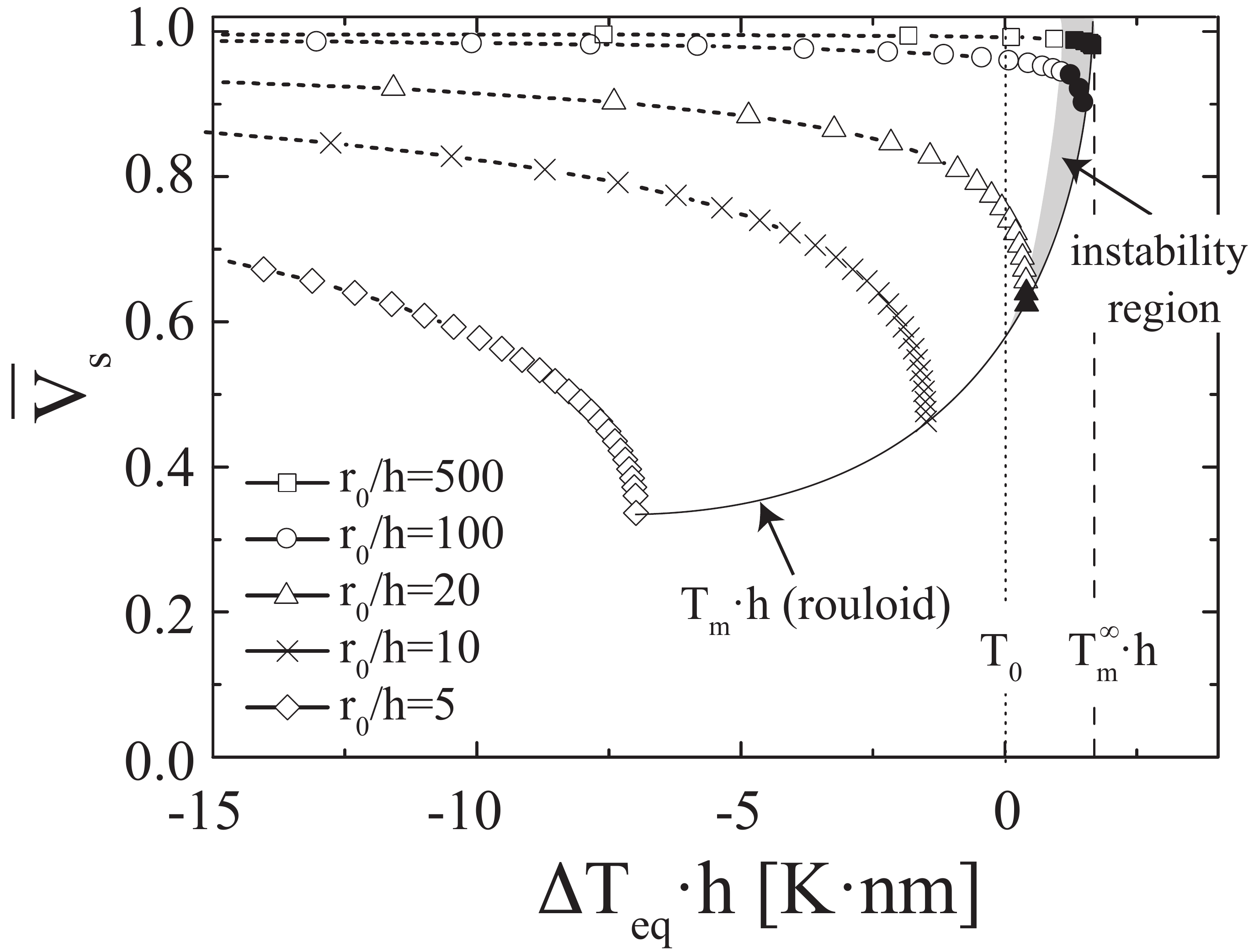}
  \caption{Melting scenarios of cylindrical islands on a planar substrate with wetting conditions as in fig.~\ref{fig:cylinders}(b).
   The data result from simulations assuming the thermodynamic data of islands of long-chain alkanes.
   The solid volume fraction $\overline{V_\text{s}}(=V_\text{s}/ \pi r^2_0 $) is plotted as function of the temperature shift $\Delta T_\text{eq}$ multiplied by the island height, $h$, for various island shapes, $r_\text{0}/h$.
   $T_0$ is the bulk melting temperature.
   $T_m^\infty \cdot h$ is the melting temperature of a straight edge terrace of height $h$.
}
  \label{fig:equil-P}
\end{figure}

Fig.~\ref{fig:equil-P} shows the solid volume fraction in the total material volume, $\overline{V_s}= V_s/ \pi r^2_0h$, as function of the melting point shift multiplied by the island height, $\overline{\Delta T} = (T_\text{m}-T_0)\cdot h$, for various island shapes, $r_\text{0}/h$.  
The endpoints of the liquid/solid coexistence regimes of each island shape denote the de facto melting temperature, $T_\text{m}(rouloid)$ multiplied by $h$.
This corresponds to the minimum possible solid volume fraction, $\overline{V_s}$, in a rouloid configuration of this island shape.
 One can see that  $T_\text{m}(rouloid)$ (solid line) covers a temperature range from below the bulk melting point (for the tall islands with small $r_\text{0}/h$) to above  bulk melting (for the low islands with large $r_\text{0}/h$).
The de facto melting point of the largest possible island i.e., a straight terrace, is denoted by $T_\text{m}^{\infty}$.
Morphological instabilities i. e., a temperature range with $T_\text{m}(bulge)<T_\text{m}(rouloid)$ with $T_\text{m}(bulge)$ on the metastable, uphill branch of  $\overline{dG_\text{I}/dV_\text{l}}$ can only occur for the low islands with the larger $r_\text{0}/h$.

The thermodynamic data used for the simulations shown in figures.~\ref{fig:island-scheme} to \ref{fig:equil-P} were taken from a real system with islands/terraces of long-chain alkanes on planar silica substrates.
For this real system, the melting behavior of multilayer islands can be observed by on-line optical microscopy for islands with a minimum radius $r_\text{0}=500\,\mathrm{nm}$ (lower optical lateral resolution limit) and heights up to $100\,\mathrm{nm}$ (the highest islands that can be prepared with radii in the range of $\mathrm{\mu m}$).
According to Fig.~\ref{fig:equil-P} for this shape with $r_\text{0}/h$ of 5 the maximum melting point shift is $\sim - 0.07$K.
This is just below the  limit of the experimental setup and indeed, no systematic melting point shifts as function of the island shape for shapes as small as $r_\text{0}/h =5$ could be measured quantitatively%%%footnote%%%
\footnote{For rather large ($r_\text{0}>>1000\,\mathrm{nm}$) and very thin islands of uniform mono-molecular thickness ($h \approx 0.4\,\mathrm{nm}$), hence $h/r_\text{0}<<1$, a pronounced melting point depression has been observed. This has been explained by a very different melting scenario than the one under consideration here (fig.~\ref{fig:island-scheme}).
In this case it is assumed that the solid monolayer islands are in thermodynamic equilibrium with a very thin (submonolayer thickness) \emph{planar} film of mobile molecules on the surrounding substrate surface.  
The thermodynamic equilibrium between the monolayer islands and the surrounding alkane submonolayer film, whose free energy is a function of its thickness, leads to the melting point depression  \cite{riegler2007pre}.}%%%
.
On the other hand, melting scenarios with rouloid and bulged morphologies are observed experimentally with a clear correlation between the morphologies and shape in agreement with the theoretical predictions.
In an earlier report, as experimental support for the theoretical analysis of the melting scenarios of straight edge terraces \cite{Halim2012bulge} we already presented some experimental observations on islands with rather large $r_\text{0}$ i. e., similar to straight edge terraces of infinite length. 
In the appendix we present additional experimental data as support for the specific melting behavior of cylinder islands analyzed in this report.

\section{Conclusion and summary}

The melting behavior of small cylindrical aggregates in a vapor environment has been analyzed in a continuum approach.
To ensure a solid cylindrical core with a certain shape i.e., a certain ratio between height and radius, specific wetting conditions were assumed:
The side wall is completely wetted by its melt while the top plane is not wetted.
The cylinder is deposited on a planar substrate, which is wetted partially by the liquid melt.
Such a system resembles the real case of islands of long chain alkanes on solid substrates.
The system behavior (liquid volume fraction, interface morphologies, melting points, etc.) is analyzed assuming material conservation (no evaporation), minimization of total energy (bulk and interfaces), and mechanical equilibrium (isobaric conditions in the liquid phase). 
It is found that depending on the cylinder geometry and the wetting conditions liquid and solid can coexist within a certain temperature range.
The system forms a solid cylindrical core surrounded by liquid with an inner (solid/liquid) and outer capillary (liquid/vapor) interface.
The capillary interface of the liquid can have a rouloid (axisymmetric) or a bulged morphology (a bulge in isobaric coexistence with a rouloid-like liquid section). 
Rouloid morphologies are possible for the entire range of liquid volume fraction and they are metastable at low liquid fractions (low temperatures).
At a certain volume fraction and above the rouloid morphology is unstable.
The system will melt without energy barrier.
The minimal unstable volume fraction corresponds to the ``de facto'' melting temperature of the rouloid morphology. 
In contrast, bulged morphologies are possible only beyond a certain threshold value of the liquid volume fraction.
Bulge morphologies are always unstable in this melting scenario.
Thus their lower limit liquid volume fraction corresponds to their ``de facto'' melting point.  
Below a certain "instability" liquid volume fraction there is an energetic barrier between isochoric rouloid and bulged morphologies.
For islands with a high ratio of radius to height this instability volume fraction lies between the bulge melting point and the higher rouloid melting point.
Hence such islands will start to melt completely by isochoric transformation of a rouloid morphology into a bulged shape and then follow a bulged melting pathway.
This will occur at a temperature above bulk melting.
Islands with a small ratio of radius to height will commence to melt completely from a rouloid morphology already below the bulk melting temperature, because the melting point of the bulged morphology has a larger liquid volume fraction than the rouloid melting point.
The calculations are in agreement with experimental observations.   
 
\section{acknowledgement}
  Discussions with H.~M\"ohwald, R. Lipowsky, H.~Kusumaatmaja and H.~Chen are gratefully acknowledged. C.~Jin was supported by IMPRS on Biomimetic Systems.

\section{Appendix}

\setcounter{figure}{0}
\setcounter{equation}{0}
\renewcommand\thefigure{A\arabic{figure}}
\renewcommand\theequation{A\arabic{equation}}

%%%%%%%%%%

\subsection{1.) The GT cylinder and the floating cylinder}

For a floating cylinder eq.~(\ref{eqn:A3}) reads as:
\begin{equation}
  \frac{dG_\text{I}^\text{cylinder}}{dV_\text{l}} = \frac{dG_\text{I}(liquid)}{dV_\text{l}} -\frac{2}{h} \gamma_\text{sv}
  \label{eqn:A9}
\end{equation} 

and thus:
\begin{align}
  \Delta T_{m}^\text{cylinder}= \frac{1}{S_\text{m}} \left[ (\frac{dG_\text{I}(liquid)}{dV_\text{l}})_\text{max}-\frac{2\gamma_\text{sv}}{h}\right]
  \label{eqn:A10}
\end{align}

For the special case of the GT approach (fig.~\ref{fig:cylinders}(a)) all energy contributions from the interfaces of the solid and liquid with the substrates can be ignored ($\gamma_\text{s1} = \gamma_\text{l1}$ and $\gamma_\text{s2} = \gamma_\text{l2}$).
In addition, because of assumption 2.) the liquid/vapour interface does not change during the phase transition. 
Therefore eq.~(\ref{eqn:A3}) reads as:
\begin{equation}
  \frac{dG_\text{I}^\text{GT}}{dV_\text{l}} = \frac{d}{dV_\text{l}}(\gamma_\text{ls}A_\text{ls})= -\frac{\gamma_\text{ls}}{r}
    \label{eqn:A5}
\end{equation} 

Thus one recovers the well-known and widely applied GT behaviour%%%footnote
\footnote{Please note that ${dG_\text{I}}/{dV_\text{l}}\propto-1/r$. Because of the minus-sign, the largest r (i. e., $r_0$) yields the largest ${dG_\text{I}}/{dV_\text{l}}$.}(citations):
\begin{equation}
  \Delta T_\text{m}^\text{GT} = \frac {\gamma_\text{ls}} {S_m}\cdot \frac{1}{r_0}
    \label{eqn:A6}
\end{equation}

In the literature this Gibbs-Thomson (GT) approach often appears  in a somewhat more general version \cite{Pawlow1909a, Pawlow1909b, Takagi1954TEM, Borel1976goldED, Couchman1977nature, Reiss1988coexist, LuKe2007rev}:

\begin{equation}
  \Delta T_\text{m}^\text{GT} (shape)= \frac{\gamma_\text{ls}}{S_\text{m}}\cdot \frac{\alpha}{r_\text{0}}
  \label{eqn:geo}
\end{equation}

Eq.~(\ref{eqn:geo}) parametrises the shape by $\alpha$, with $\alpha = 1$ for cylindrical and $\alpha = 2$ for spherical geometries \cite{unruh1993melting}).
As shown by Eq.~(\ref{eqn:A6}) this approach is only correct, if energy contributions from the cylinder base planes can be neglected (fig.~\ref{fig:cylinders}(a)). 
This is only the case for a few simple configurations, such as freely suspended spheres or long cylinders\cite{nozieres1992shape}.

The GT model extended to cases with anisotropic solid/liquid interfacial energies (asymmetric Wulff polyhedra) leads to the Gibbs-Thomson-Herring (GTH) approach \cite{nozieres1992shape}.
The GTH approach may be applied to (small) solid aggregates embedded in its infinite-size liquid (melt) or infinite-size vapour phase.
However, it is not a reasonable description of small solid aggregates embedded in a finite size melt.
Both, the GT and the GTH approach do \emph{not} consider the impact of the capillary interface. 

%%%%%%%%%%%%%%%%%%%%%

\subsection{2.) Experimental data showing rouloid and bulge morphologies}

\begin{figure}
  \centering
  \includegraphics[width=\columnwidth]{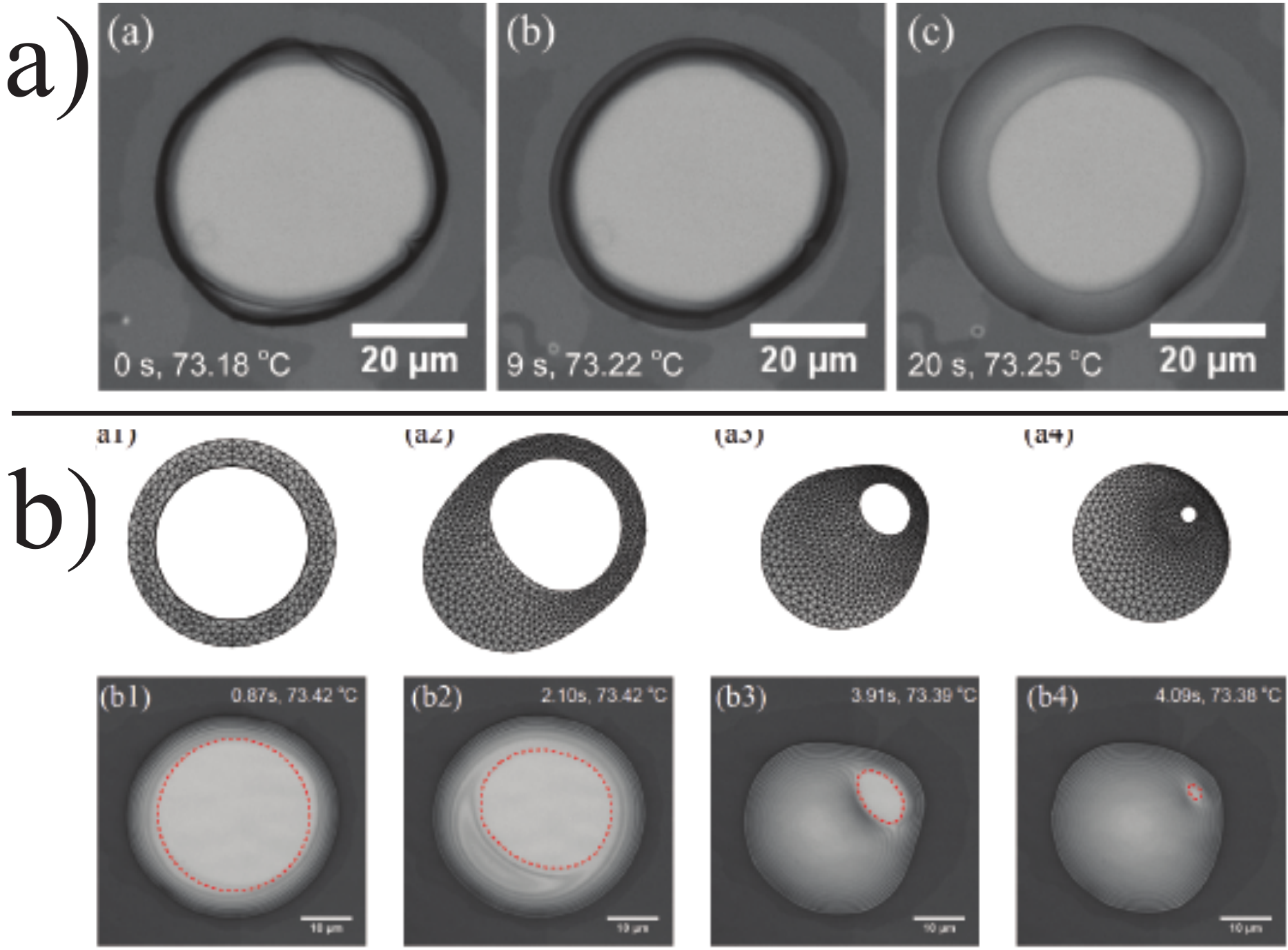}
  \caption{(a) Evolution of a rouloid liquid channel morphology as observed for alkane islands ($C_{36}H_{74}$, $r_0\approx 50\mu m, h\approx 2.2\mu m$, $\Theta \approx 20 ^\circ$) during a slight temperature increase.
  (b) Simulated (top row) and observed (lower row) evolution of a bulged liquid channel morphology as observed for alkane islands under approximately isothermal conditions (Experimental data: $C_{36}H_{74}$, $r_0\approx 18\mu m, h\approx 1\mu m$, $\Theta \approx 20 ^\circ$, Simulation data: $r_0/h$=20, $\Theta =15^\circ$,  
  } 
  \label{fig:exp-results}
\end{figure}

Fig.~\ref{fig:exp-results} shows experimental data on the melting scenario of cylinder islands of long chain alkane islands.
Fig.~\ref{fig:exp-results}(a) reveals how the rouloid volume increases as the temperature is increasing slowly. 
Fig.~\ref{fig:exp-results}(b) shows how a rouloid morphology transforms (isothermally) into a bulged morphology while the liquid volume fraction increases.
 The top row of the Figure shows the simulation of the evolution of the surface morphology,
 the lower row experimental observations for similar real shapes/conditions. 
 The simulation data are resized to the experimental scales and the red lines in the experimental data show the residual solid cylinder core from the simulated data projected on the experimental observations.

\bibliography{meltingref.bib}

\end{document}